\documentclass[12pt]{article}
\usepackage{geometry}
\usepackage{graphicx}
\usepackage{amsmath,amssymb}
\usepackage{cite}
\newcommand{\lw}[1]{\smash{\lower 1.5ex\hbox{#1}}}
\newcommand{\ri}[1]{\smash{\raise 1.5ex\hbox{#1}}}
\newcommand{\riw}[1]{\smash{\raise 3.0ex\hbox{#1}}}

\begin{document}

\title{Unified Description of Multiplicity Distributions and Bose-Einstein Correlations at the LHC Based on the Three-Negative Binomial Distribution}

\author{Minoru Biyajima$^{1}$ and Takuya Mizoguchi$^{2}$\\
{\small $^{1}$Department of Physics, Shinshu University, Matsumoto 390-8621, Japan}\\
{\small $^{2}$National Institute of Technology, Toba College, Toba 517-8501, Japan}}

\maketitle

\begin{abstract}
Using the Monte Carlo data at 7 TeV collected by the ATLAS collaboration (PYTHIA 6), we examine the necessity of applying the three-negative binomial distribution (T-NBD). By making use of the T-NBD formulation, we analyze the multiplicity distribution (MD) and the Bose-Einstein correlation (BEC) at the Large Hadron Collider (LHC). In the T-NBD framework, the BEC is expressed by two degrees of coherence $\lambda_1$ and $\lambda_2$ and two kinds of exchange functions $E_1^2$ and $E_2^2$ that act over interaction ranges $R_1$ and $R_2$, respectively. Using the calculated $\lambda_1$ and $\lambda_2$ based on the T-NBD, along with free $\lambda_1$ and $\lambda_2$, we analyze the BEC data at 0.9 and 7 TeV. The estimated parameters $R_1$ and $R_2$ are almost coincident and seem to be consistent with $pp$ collisions. We also present an enlarged Koba-Nielsen-Olesen (KNO) scaling function based on the T-NBD, and apply it to KNO scaling at LHC energies. The enlarged scaling function describes the observed violation of the KNO scaling.
\end{abstract}


\section{\label{sec1}Introduction}


\subsection{Negative Binomial Distribution (NBD)}

Approximately three decades ago, the UA5 collaboration~\cite{Fuglesang:1989st} discovered a violation of Koba-Nielsen-Olesen (KNO) scaling~\cite{Koba:1972ng} at CERN S$\bar pp$S collider. To explain those data, the UA5 collaboration assumed a double-negative binomial distribution (D-NBD) of the data. The NBD is given as
\begin{eqnarray}
 P_{\rm NBD}(n,\,k,\,\langle n\rangle) = \frac{\Gamma (n+k)}{\Gamma (n+1)\Gamma (k)}\frac{(\langle n\rangle/k)^n}{(1+\langle n\rangle/k)^{n+k}},
\label{eq1}
\end{eqnarray}
where $\langle n\rangle$ and $k$ are the averaged multiplicity and the intrinsic parameter of the NBD, respectively. The D-NBD is expressed as
\begin{eqnarray}
  P_{\rm (D\mathchar`-N)}(n,\, \langle n\rangle,\, \alpha_i,\, k_i) = \sum_{i=1}^2 \alpha_i P_{{\rm NBD}_i}(n,\,\langle n_i\rangle,\,k_i),
\label{eq2}
\end{eqnarray}
where $\alpha_1 + \alpha_2 = 1$. Using Eqs.~(\ref{eq1}) and (\ref{eq2}), several authors~\cite{Giovannini:1998zb,Ghosh:2012xh,Zaccolo:2015udc,Giovannini:2003ft} have analyzed the multiplicity distribution (MD) of data at high energies, such as the high-energy data of the Large Hadron Collider (LHC)~\cite{Aad:2010ac,Khachatryan:2010nk}. Among those analyses, the three-NBD (T-NBD) was applied by Zborovsky~\cite{Zborovsky:2013tla}:
\begin{eqnarray}
  P_{\rm (T\mathchar`-N)}(n,\, \langle n\rangle,\, \alpha_i,\, k_i) = \sum_{i=1}^3 \alpha_i P_{{\rm NBD}_i}(n,\,\langle n_i\rangle,\,k_i).
\label{eq3}
\end{eqnarray}
where $\alpha_1 + \alpha_2 + \alpha_3 = 1.0$ (see also \cite{Giovannini:2003ft}).\medskip

Here, we address the question ``Why must the MD of LHC data be analyzed in the T-NBD framework?'' To find a plausible answer, we must consider the constraints adopted by the ATLAS and CMS collaborations (for example, $|\eta|<2.5$ and $|\eta|<2.4$). Even in restricted $\eta$ regions, Monte Carlo (MC) estimations have revealed that three processes, i.e., the non-diffractive dissociation (ND), the single-diffractive dissociation (SD), and the double-diffractive dissociation (DD) contribute to the MD at LHC energies~\cite{ATLAS:2010mza,GrosseOetringhaus:2009kz,Navin:2010kk}.

For the reader's convenience, our analysis results of MC data at 7 TeV collected by the ATLAS collaboration are presented in Tables~\ref{tab11} and \ref{tab12}, and Figs.~\ref{fig5} and \ref{fig6} of Appendix \ref{secA}. The individual data ensembles of ND, SD, and DD, defined by the partial probability distributions $P_{\rm ND}(n)$, $P_{\rm SD}(n)$, and $P_{\rm DD}(n)$ respectively, are reasonably described by the D-NBD (see Table \ref{tab1}). Accordingly, the total probability $P_{\rm tot} = P_{\rm ND} + P_{\rm SD} + P_{\rm DD}$ is also described by the T-NBD. This behavior implies an important role for the T-NBD in MD analyses at LHC energies. 


\begin{table}[htbp]
\centering
\caption{\label{tab1}Stochastic properties of the three ensembles computed by PYTHIA 8 in the ATLAS experiment at 7 TeV (see Figs.~\ref{fig5} and \ref{fig6} in Appendix~\ref{secA}).}
\vspace{1mm}
\begin{tabular}{c|c|c|c}
\hline
QCD      & acceptance and       & \lw{stochastic property}    & \lw{stochastic description}\\
PYTHIA 6 & correction by & \lw{of individual ensemble} & \lw{of sum of three ensembles}\\
7 TeV    & ATLAS coll.~\cite{ATLAS:2010mza}   &                        & \\
\hline
\lw{ND} & $P_{\rm ND}(n)$ & (sum of) & \\
   & $f_{\rm ND} = \sum P_{\rm ND}(n)= 0.787$ & NBD$_1$ and NBD$_2$ & T-NBD  Eq.~(\ref{eq3})\\
\cline{1-3}
\lw{SD} & $P_{\rm SD}(n)$ &  & \lw{$P_{\rm (T\mathchar`-N)}(n,\, \langle n\rangle,\, \alpha_i,\, k_i) =\hspace{5mm}$}\\
   & $f_{\rm SD} = 0.121$ & NBD$_2$ and NBD$_3$ & \lw{$\sum_{i=1}^3 \alpha_i P_{{\rm NBD}_i}(n,\,\langle n_i\rangle,\,k_i)$}\\
\cline{1-3}
\lw{DD} & $P_{\rm DD}(n)$ &  & \\
   & $f_{\rm DD} = 0.092$ & NBD$_3$ and NBD$_2$ & \\
\hline
   & sum of fractions: & NBD$_i$'s are specified & Eq.~(\ref{eq3}) is a \\
   & $f_{\rm ND}+f_{\rm SD}+f_{\rm DD} = 1.0$ & by $(\langle n\rangle,\,k)$ & possible candidate\\
\hline
\end{tabular}
\end{table}

Very recently, Zborovsky~\cite{Zborovsky:2018vyh} revealed the stochastic structure of the T-NBD studying the oscillations in combinations of T-NBDs. As pointed out by Wilk and Wlodarczyk~\cite{Wilka:2016ufh}, Zborovsky's work supports the theoretical plausibility of the T-NBD. See also recent study on this subject~\cite{Wilk:2018fhw}. Regarding recent T-NBD investigations, we approach the T-NBD from a different perspective, namely, the identical particle effect~\cite{Biyajima:1978cz,Biyajima:1982un,Biyajima:1990ku} observed at the LHC (see Refs.~\cite{Mizoguchi:2010vc,Biyajima:2018abe,Mizoguchi:2019cra} and \cite{Aad:2015sja,Khachatryan:2011hi,Aaij:2017oqu} for related theoretical and empirical studies, respectively)\footnote{In Ref.~\cite{Biyajima:2018abe}, for $N^{\rm BG}$, an identical separation between two ensembles with $\alpha_1$and $\alpha_2$ is assumed. For no-separation between them, the following formula is obtained:
$$
N^{\rm (2+:\,2-)}/N^{\rm BG} = 1 + (a_1/s)E_1^2 +(a_2/s)E_2^2,
$$
where $s=a_1+a_2 =\alpha_1\langle n_1\rangle^2 + \alpha_2\langle n_2\rangle^2$ (see succeeding Ref.~\cite{Mizoguchi:2019cra}).
}.


\subsection{Bose-Einstein correlation at the LHC}

The moments of a charged-particle distributions are calculated as 
\begin{eqnarray}
  \langle n\rangle &=& \sum_{i=1}^3 \alpha_i \langle n_i\rangle,\nonumber\\
  \langle n(n-1)\rangle &=& \sum_{i=1}^3 \alpha_i \langle n_i(n_i-1)\rangle 
  = \sum_{i=1}^3 \alpha_i \langle n_i\rangle^2 \left(1.0 + \frac{1.0}{k_i}\right),\nonumber\\
  \langle n(n-1)(n-2\rangle &=& \sum_{i=1}^3 \alpha_i \langle n_i(n_i-1)(n_i-2)\rangle 
  = \sum_{i=1}^3 \alpha_i \langle n_i\rangle^3 \left(1.0 + \frac{3.0}{k_i} + \frac{2.0}{k_i^2}\right).
\label{eq4}
\end{eqnarray}
When the particles are identical, we obtain the following relation (where the charge sign $a$ is $+$ or $-$):
\begin{eqnarray}
  \langle n^a(n^a-1)\rangle = \sum_{i=1}^3 \alpha_i \langle n_i^a\rangle^2 \left(1.0 + \frac{2.0}{k_i}\right),
\label{eq5}
\end{eqnarray}
Eq.~(\ref{eq5}) can be interpreted as
\begin{eqnarray*}
{\rm Eq.\ }(\ref{eq5}):&& \sum_{i=1}^3 (\mbox{The number of pairs of identical charged particle in MD($P(n)$) with $\alpha_i$})\\
&& \qquad\times (\mbox{identical particle effect in MD}).
\end{eqnarray*}

Meanwhile, the authors of \cite{Biyajima:2018abe} recently studied the interrelation between the MD and the Bose-Einstein correlation (BEC) under the D-NBD assumption. To extend the framework of T-NBD, we compute $N^{\rm BG}$ as a function of $\{\alpha_i,\,\langle n_i\rangle:\, i=1\sim 3\}$ (see Ref.~\cite{Mizoguchi:2019cra}). The BEC in the extended framework is given by
\begin{eqnarray}
 \frac{N^{\rm (2+:\,2-)}}{N^{\rm BG}} 
=\dfrac{\sum_{i=1}^3 \alpha_i (\langle n_i^+\rangle^2+\langle n_i^-\rangle^2) \left(1.0 + \frac{2.0}{k_i}\right)}{\sum_{i=0}^3 2\times(\mbox{The number of pairs of different charged particles $(+-)$ in MD with $\alpha_i$})}.
\label{eq6}
\end{eqnarray}
To simplify the calculations, we denote $\langle n_i^+\rangle = \langle n_i^-\rangle = \langle n_i\rangle/2$. In the denominator $N^{\rm BG}$ of Eq.~(\ref{eq6}), the three coefficients $\alpha_i \langle n_i\rangle^2$ are given by
\begin{eqnarray}
  N^{\rm BG} &=& \sum_{i=1}^3 \alpha_i \langle n_i\rangle^2 = a_1+a_2+a_3=s,\nonumber\\
  &&\sum_{i=1}^3\left(\frac{a_i}s\right)=1.
\label{eq7}
\end{eqnarray}
In the framework of the T-NBD assumption, we have
\begin{eqnarray}
 \frac{N^{\rm (2+:\,2-)}}{N^{\rm BG}} 
= \sum_{i=1}^3\left(\frac{a_i}s\right)\left(1.0+\frac 2{k_i}E_{{\rm BEC}_i}^2\right)
\label{eq8}
\end{eqnarray}
To describe the BEC in the $0\le Q\le 2$ GeV region, we assume the following exchange function $E_{\rm BEC}^2$
\begin{eqnarray}
E_{\rm BEC}^2 =\ \left\{
\begin{array}{l}
\exp(-RQ) \mbox{ (Exponential function) (E)},\medskip\\
\exp(-(RQ)^2) \mbox{ (Gaussian distribution) (G)},
\end{array}
\right.
\label{eq9}
\end{eqnarray}
where $R$ and $Q$ are the interaction range and the momentum-transfer squared function, respectively. The latter is calculated as $Q= \sqrt{-(p_1-p_2)^2}$, where $p_1$ and $p_2$ are momenta of identical particles.

By making use of those calculations mentioned above, the BEC is then formulated as
\begin{eqnarray}
 {\rm BEC}_{\rm (T\mathchar`-N)} 
&=& 1.0 + \sum_{i=1}^3\left(\frac{a_i}s\right)\left(\frac 2{k_i}\right)E_{{\rm BEC}_i}^2\nonumber\\
&=& 1.0 + \lambda_1^{\rm (T\mathchar`-N)} E_{\rm BEC_1}^2
    + \lambda_2^{\rm (T\mathchar`-N)} E_{\rm BEC_2}^2
    + \mathcal{O}(10^{-3}),
\label{eq10}
\end{eqnarray}
where $\lambda_i^{\rm (T\mathchar`-N)} = (a_i/s)(2/k_i)$ ($i=1,\,2$). It should be noted that the third component (with coefficient $\alpha_3$) exhibits a Poisson property. Because the $k_3$ values are large, the third term ($i=3$) does not numerically contribute to BEC$_{\rm (T\mathchar`-N)}$.

In our BEC analysis of LHC data, we note that all three collaborations (ATLAS, CMS, and LHCb~\cite{Aad:2015sja,Khachatryan:2011hi,Aaij:2017oqu}) applied the well-known conventional formula
\begin{eqnarray}
{\rm CF_I} = 1.0 + \lambda E_{\rm BEC}^2,
\label{eq11}
\end{eqnarray}
Regarding Eq.~(\ref{eq10}) as another conventional formula, we would like to propose that
\begin{eqnarray}
{\rm CF_{II}} = 1.0 + \lambda_1^{\rm (II)} E_{\rm BEC_1}^2+\lambda_2^{\rm (II)} E_{\rm BEC_2}^2,
\label{eq12}
\end{eqnarray}
where $\lambda_1^{\rm (II)}$ and $\lambda_2^{\rm (II)}$ are free parameters. By analyzing the BEC data, we can compare $\lambda_1^{\rm (II)}$ and $\lambda_2^{\rm (II)}$ in Eq.~(\ref{eq12}) with the terms $\lambda_i^{\rm (T\mathchar`-N)} = (a_i/s)(2/k_i)$ ($i=1,\,2$) in Eq.~(\ref{eq10}), and the terms $R_i^{\rm (T\mathchar`-N)}$ and $R_i^{\rm (II)}$ in Eqs.~(\ref{eq10}) and (\ref{eq12}).\medskip\\

The second section of this paper analyzes the MD at 0.9 and 7 TeV through Eq.~(\ref{eq3}). The second and third moments, and $a_i$'s and $(a_i/s)(2/k_i)$'s are displayed in this section. Section \ref{sec3} analyzes the BEC through Eqs.~(\ref{eq10}), (\ref{eq11}) and (\ref{eq12}). This paper concludes with remarks and discussions in Section \ref{sec4}. The appendices analyze the MC data at 7 TeV collected by the ATLAS collaboration (Appendix~\ref{secA}) and the KNO scaling data by an enlarged KNO scaling function based on the T-NBD assumption (Appendix \ref{secB}).


\section{\label{sec2} Multiplicity Distribution ($P(n)$) Analysis} 
\begin{figure}[htbp]
  \centering
  \includegraphics[width=0.48\columnwidth]{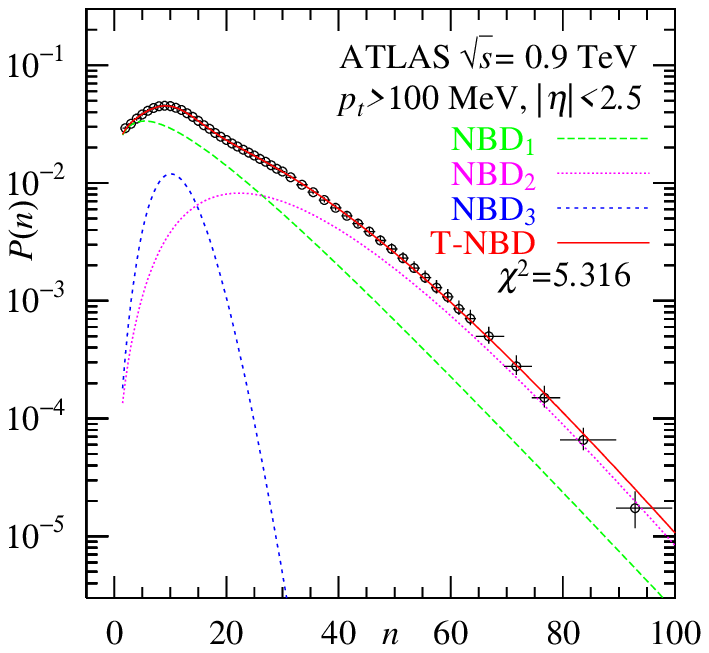}
  \includegraphics[width=0.48\columnwidth]{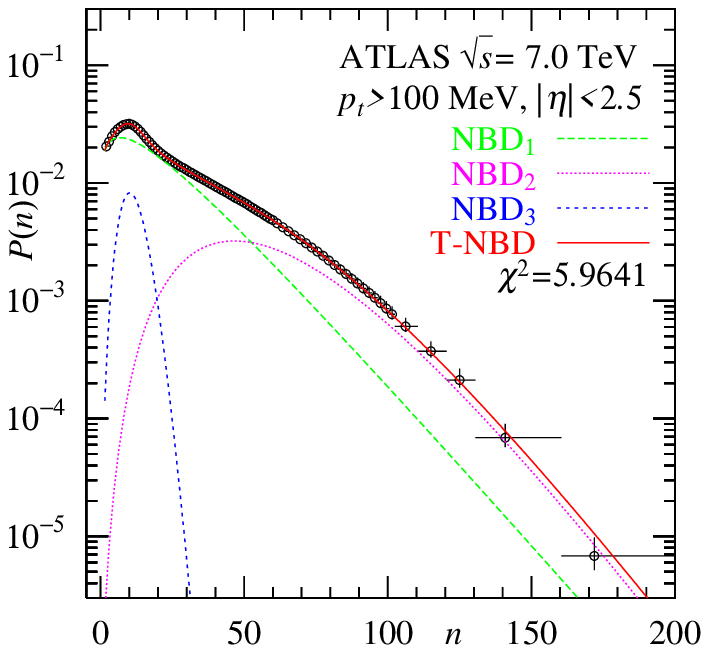}\\
  \includegraphics[width=0.48\columnwidth]{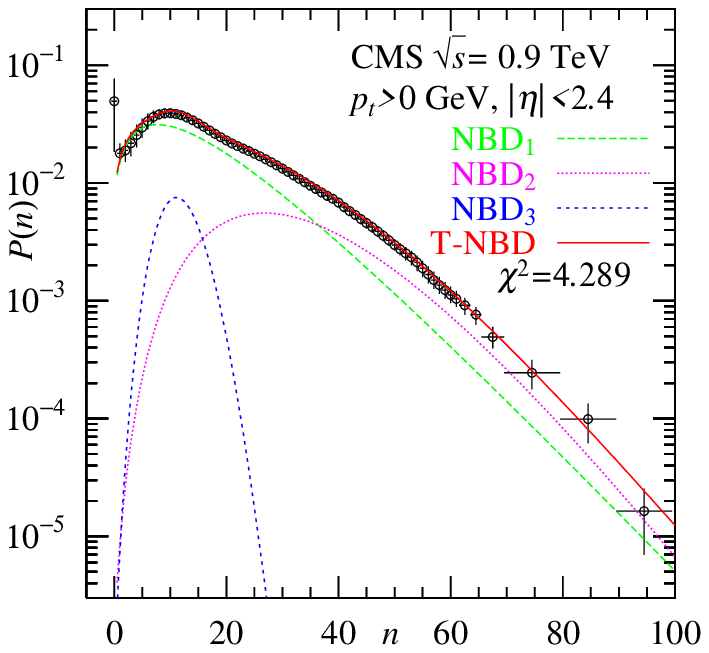}
  \includegraphics[width=0.48\columnwidth]{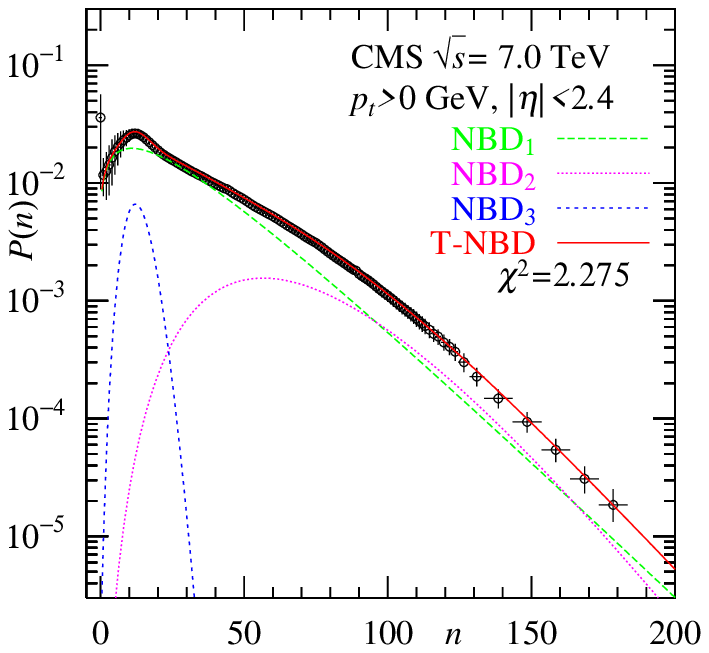}
  \caption{\label{fig1}Analysis of MD ($P(n)$) data collected by the ATLAS and CMS collaborations. The $P(n)$ data were computed by Eq.~(\ref{eq3}). As done in Ref.~\cite{Zborovsky:2013tla}, we exclude $P(0)$ obtained by the CMS in our analysis. All data are renormalized.}
\end{figure}

We begin by analyzing the MD at 0.9 TeV and 7 TeV obtained by the ATLAS~\cite{Aad:2010ac} and CMS~\cite{Khachatryan:2010nk} collaborations under the T-NBD assumption. The MINUIT program is initialized by assigning random variables to the physical quantities. The estimated parameters are displayed in Fig.~\ref{fig1} and Table~\ref{tab2}. Note that both collaborations obtained similar minimum $\chi^2$ values at 0.9 TeV. To compare our results with those of Zborovsky~\cite{Zborovsky:2013tla}, we adopt the same treatments to the probability distributions. Specifically, we renormalize Eq.~(\ref{eq3}) without the $P(0)$ and $P(1)$ as the MD obtained by the ATLAS collaboration, and also renormalize the MD obtained by the CMS collaboration after excluding $P(0)$.


\begin{table}[htbp]
\centering
\caption{\label{tab2}Analysis (by Eq.~(\ref{eq3})) of the MD ($P(n)$) data at 0.9 and 7.0 TeV collected by the ATLAS and CMS collaborations.} 
\vspace{1mm}
\begin{tabular}{c|ccccc}
\hline
& $i$ & $\alpha_i$ & $\langle n_i\rangle$ & $a_i$ & $k_i$\\
\hline
ATLAS
& 1
& 0.640$\pm$0.199
& 13.493$\pm$2.546
& 116.519$\pm$56.975
& 1.78$\pm$0.20\\
0.9 TeV
& 2
& 0.250$\pm$0.164
& 28.488$\pm$3.588
& 202.892$\pm$142.572
& 5.01$\pm$1.37\\
$\chi^2 = 5.317$
& 3
& 0.111$\pm$0.047
& 10.998$\pm$0.237
& 13.426$\pm$5.714
& 28.1$\pm$24.4\\
\hline
ATLAS
& 1
& 0.737$\pm$0.053
& 21.934$\pm$2.392
& 354.571$\pm$81.430
& 1.50$\pm$0.08\\
7.0 TeV
& 2
& 0.183$\pm$0.061
& 57.214$\pm$2.597
& 599.040$\pm$206.953
& 5.67$\pm$0.76\\
$\chi^2 = 5.964$
& 3
& 0.080$\pm$0.010
& 11.164$\pm$0.169
& 9.971$\pm$1.282
& 23.4$\pm$8.4\\
\hline
ATLAS
& 1
& 0.754$\pm$0.063
& 22.625$\pm$2.549
& 385.966$\pm$92.755
& 1.48$\pm$0.08\\
7.0 TeV
& 2
& 0.164$\pm$0.063
& 57.936$\pm$2.618
& 550.479$\pm$217.238
& 5.94$\pm$0.98\\
$\chi^2 = 6.160$
& 3
& 0.082$\pm$0.010
& 11.177$\pm$0.178
& 10.244$\pm$1.291
& 23.4$\pm$8.7\\
\hline
CMS
& 1
& 0.743$\pm$0.179
& 15.852$\pm$2.454
& 186.705$\pm$73.245
& 2.08$\pm$0.20\\
0.9 TeV
& 2
& 0.189$\pm$0.170
& 32.160$\pm$4.567
& 195.476$\pm$184.382
& 6.56$\pm$2.85\\
$\chi^2 = 4.289$
& 3
& 0.068$\pm$0.032 
& 11.624$\pm$0.814
& 9.188$\pm$4.511
& 896$\pm$817\\
\hline
CMS
& 1
& 0.739$\pm$0.200
& 15.830$\pm$2.841
& 185.092$\pm$83.177
& 2.10$\pm$0.21\\
0.9 TeV
& 2
& 0.193$\pm$0.179
& 32.093$\pm$5.002
& 199.236$\pm$194.669
& 6.49$\pm$3.03\\
$\chi^2 = 4.941$
& 3
& 0.068$\pm$0.031
& 11.618$\pm$0.810
& 9.177$\pm$4.384
& $\infty$\\
\hline
CMS
& 1
& 0.826$\pm$0.091
& 28.613$\pm$4.126
& 676.108$\pm$208.928
& 1.66$\pm$0.12\\
7.0 TeV
& 2
& 0.103$\pm$0.098
& 67.206$\pm$6.727
& 465.270$\pm$452.384
& 6.67$\pm$2.92\\
$\chi^2 = 2.275$
& 3
& 0.071$\pm$0.028
& 13.018$\pm$0.870
& 12.036$\pm$5.012
& 38.1$\pm$73.6\\
\hline
\end{tabular}
\end{table}


\begin{table}[htbp]
\centering
\caption{\label{tab3}Second and third moments of MD at 0.9 TeV and 7 TeV, measured (data) and calculated by Eqs.~(\ref{eq2}) (D-NBD) and (\ref{eq3}) (T-NBD).} 
\vspace{1mm}
\begin{tabular}{|c|ccc|ccc|}
\hline
 & \multicolumn{3}{c|}{$\langle n(n-1)\rangle$\quad ($\times 10^3$)} & \multicolumn{3}{c|}{$\langle n(n-1)(n-2)\rangle$\quad ($\times 10^4$)}\\
 & data & D-NBD & T-NBD & data & D-NBD & T-NBD\\
\hline
ATLAS 0.9 TeV 
& 0.454$\pm$0.026
& 0.450
& 0.439
& 1.55$\pm$0.13
& 1.54
& 1.53\\
ATLAS 7.0 TeV   
& 1.35$\pm$0.08
& 1.35
& 1.30
& 8.75$\pm$0.79
& 9.09
& 8.43\\
\hline
CMS 0.9 TeV
& 0.488$\pm$0.050
& 0.525
& 0.511
& 1.73$\pm$0.22
& 1.86
& 1.82\\
CMS 7.0 TeV
& 1.57$\pm$0.13
& 1.67
& 1.63
& 10.94$\pm$1.10
& 11.89
& 11.52\\
\hline
\end{tabular}
\end{table}

Our results in Table \ref{tab2} almost match those of Zborovsky~\cite{Zborovsky:2013tla}. Moreover, the empirical values of the second and third moments almost equal those of the D-NBD and T-NBD (Table~\ref{tab3}). From the values in Table \ref{tab2}, we obtain $\lambda_i^{\rm (T\mathchar`-N)} = (\alpha_i \langle n_i\rangle^2/s)(2/k_i)$ ($i=1,\,2$). The results are shown in Table~\ref{tab4}.


\begin{table}[htbp]
\centering
\caption{\label{tab4}Values of $\lambda_i^{\rm (T\mathchar`-N)} = (\alpha_i \langle n_i\rangle^2/s)(2/k_i)$.} 
\vspace{1mm}
\begin{tabular}{c|ccc}
\hline
& $\lambda_1^{\rm (T\mathchar`-N)}$ & $\lambda_2^{\rm (T\mathchar`-N)}$ & $\lambda_3^{\rm (T\mathchar`-N)}$\\
\hline
ATLAS 0.9 TeV, $\chi^2 = 5.317$
& 0.393$\pm$0.214
& 0.244$\pm$0.103\\

ATLAS 7.0 TeV, $\chi^2 = 5.964$
& 0.490$\pm$0.130
& 0.219$\pm$0.045\\

ATLAS 7.0 TeV, $\chi^2 = 6.160$
& 0.552$\pm$0.152
& 0.196$\pm$0.050
& $\mathcal{O}(10^{-3}\sim 10^{-4})$\\

CMS 0.9 TeV, $\chi^2 = 4.289$
& 0.460$\pm$0.241
& 0.152$\pm$0.102\\

CMS 0.9 TeV, $\chi^2 = 4.941$
& 0.448$\pm$0.250
& 0.156$\pm$0.110\\

CMS 7.0 TeV, $\chi^2 = 2.275$
& 0.706$\pm$0.296
& 0.121$\pm$0.091\\
\hline
\end{tabular}
\end{table}


\section{\label{sec3}Analyses of BEC data by Eqs.~(\ref{eq10}), (\ref{eq11}) and (\ref{eq12})}
\begin{figure}[htbp]
  \centering
  \includegraphics[width=0.48\columnwidth]{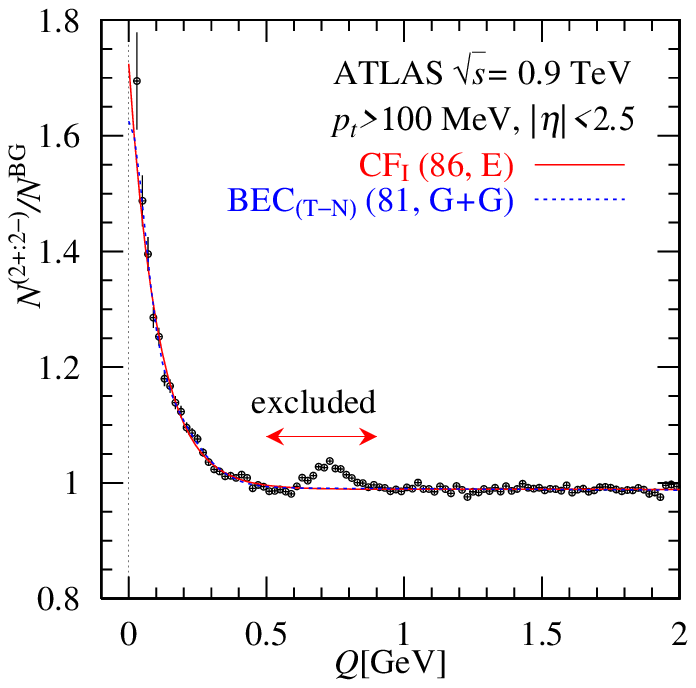}
  \includegraphics[width=0.48\columnwidth]{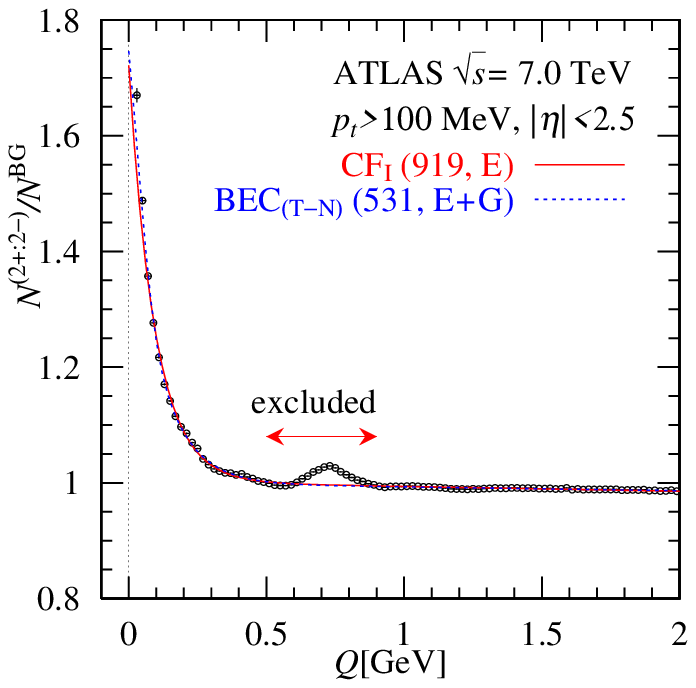}\\
  \includegraphics[width=0.48\columnwidth]{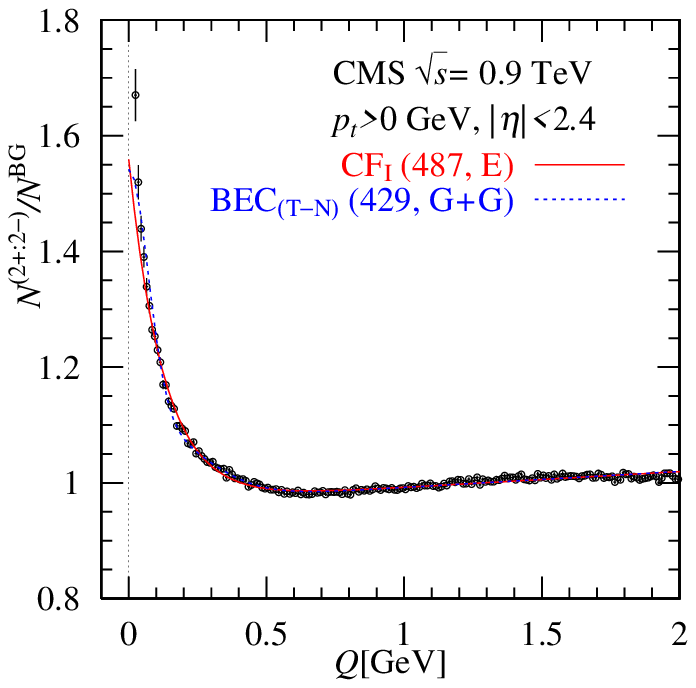}
  \includegraphics[width=0.48\columnwidth]{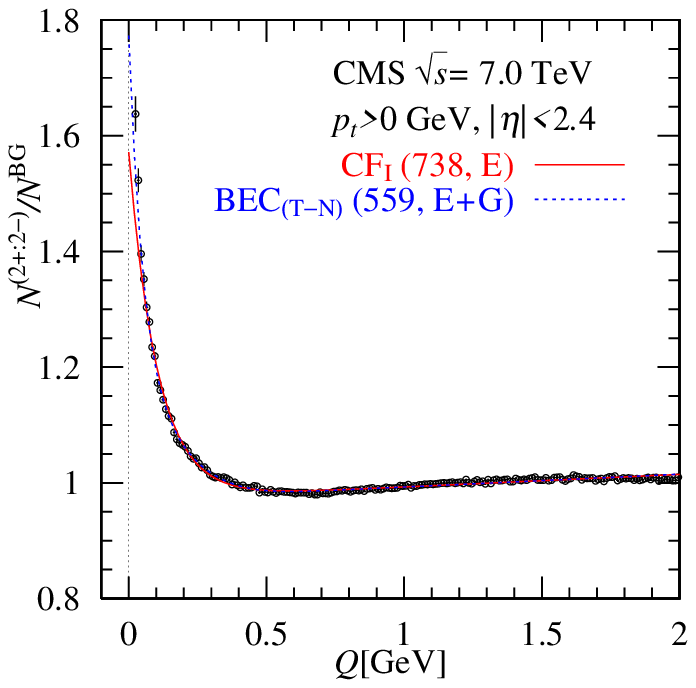}
  \caption{\label{fig2} BEC data at 0.9 and 7 TeV, collected by the ATLAS and CMS collaborations and analyzed by Eqs.~(\ref{eq10}) and (\ref{eq11}). Values in parentheses are the $\chi^2$ values in different geometrical combinations of the exponential function (E) and Gaussian distribution (G).}
\end{figure}

Our BEC results are displayed in Fig.~\ref{fig2} and Tables~\ref{tab5} and \ref{tab6}. In Tables~\ref{tab5} and \ref{tab6}, the combinations exhibiting high coincidence are indicated by ($\ast 1$ and $\ast 2$). In the BEC$_{\rm (T\mathchar`-N)}$ analysis, we apply the calculated $\lambda_i^{\rm (T\mathchar`-N)}$ ($i=1,\,2$) values in Table \ref{tab4}, which were fixed in the MINUIT computations. Contrarily, the four parameters of the CF$_{\rm II}$, calculations \{$R_i^{\rm (II)}$, $\lambda_1^{\rm (II)}$, $i=1,\,2$\} are free; however, the four parameters in Eq.~(\ref{eq12}) and the set of two parameters ($R_i^{\rm (T\mathchar`-N)}$, $i=1,\,2$) and two fixed parameters ($\lambda_i^{\rm (T\mathchar`-N)}$, $i=1,\,2$) give very similar results. We emphasize that the geometrical combinations G$+$G at 0.9 TeV and E$+$G at 7 TeV are identical in the CF$_{\rm II}$ and BEC$_{\rm (T\mathchar`-N)}$ formulations. This coincidence is likely attributable to the common stochastic properties of the MD and BEC ensembles (see Tables \ref{tab5} and \ref{tab6}). 


\begin{table}[htbp]
\vspace{-5mm}
\centering
\caption{\label{tab5}Analysis results of the BEC data from ATLAS~\cite{Aad:2015sja} using Eqs.~(\ref{eq10}), (\ref{eq11}), and (\ref{eq12}), where the BEC formulas are normalized with a consistent factor and the long-range correlation is assumed as $(1+\varepsilon Q)$ (the labels ($\ast 1$ and $\ast 2$) indicate equivalence between the tabulated results by Eq.~(\ref{eq10}) and Eq.~(\ref{eq12}), respectively).}
\vspace{1mm}
\begin{tabular}{c|ccccc}
\multicolumn{2}{l}{ATLAS\ 0.9 TeV}\\
\hline
& $R$ [fm] & $\lambda$ (free) & --- & --- & $\chi^2/$ndf\\
\cline{2-6}
CF$_{\rm I}$
& 1.84$\pm$0.07 (E)
& 0.74$\pm$0.03
& ---
& ---
& 86.0/75\\
& 1.00$\pm$0.03 (G)
& 0.34$\pm$0.01
& ---
& ---
& 148/75\\
\hline
& $R_1^{\rm (II)}$ [fm] & $\lambda_1^{\rm (II)}$ (free) & $R_2^{\rm (II)}$ [fm] & $\lambda_2^{\rm (II)}$ (free) & $\chi^2$\\
\cline{2-6}
CF$_{\rm II}$
& 4.52$\pm$1.02 (E)
& 0.98$\pm$0.21
& 0.81$\pm$0.05 (G)
& 0.21$\pm$0.04
& 78.2\\
& 2.82$\pm$0.28(G)
& 0.47$\pm$0.07
& 0.87$\pm$0.03 (G)
& 0.26$\pm$0.02
& 79.8 ($\ast 1$)\\
\hline
BEC$_{\rm (T\mathchar`-N)}$ & $R_1^{\rm (T\mathchar`-N)}$ [fm] & $\lambda_1^{\rm (T\mathchar`-N)}$ (calcu.) & $R_2$ [fm] & $\lambda_2^{\rm (T\mathchar`-N)}$ (calcu.) & $\chi^2$\\
\cline{2-6}
MD $\chi^2=5.32$
& 2.55$\pm$0.10 (G)
& 0.39
& 0.85$\pm$0.02 (G)
& 0.25
& 81.1 ($\ast 1$)\\
& 3.37$\pm$0.22 (E)
& 0.39
& 0.89$\pm$0.02 (G)
& 0.24
& 101\\
\hline
\multicolumn{2}{l}{ATLAS 7.0 TeV}\\
\hline
& $R$ [fm] & $\lambda$ (free) & --- & --- & $\chi^2/$ndf\\
\cline{2-6}
CF$_{\rm I}$
& 2.06$\pm$0.01 (E)
& 0.72$\pm$0.01
& ---
& ---
& 919/75\\
& 1.13$\pm$0.01 (G)
& 0.33$\pm$0.00
& ---
& ---
& 4578/75\\
\hline
& $R_1^{\rm (II)}$ [fm] & $\lambda_1^{\rm (II)}$ (free) & $R_2^{\rm (II)}$ [fm] & $\lambda_2^{\rm (II)}$ (free) & $\chi^2$\\
\cline{2-6}
CF$_{\rm II}$
& 6.54$\pm$0.40 (E)
& 0.73$\pm$0.05
& 1.80$\pm$0.02 (E)
& 0.54$\pm$0.02
& 465\\
& 1.85$\pm$0.02 (E)
& 0.59$\pm$0.01
& 3.51$\pm$0.12 (G)
& 0.28$\pm$0.01
& 466 ($\ast 2$)\\
\hline
BEC$_{\rm (T\mathchar`-N)}$ & $R_1^{\rm (T\mathchar`-N)}$ [fm] & $\lambda_1^{\rm (T\mathchar`-N)}$ (calcu.) & $R_2^{\rm (T\mathchar`-N)}$ [fm] & $\lambda_2^{\rm (T\mathchar`-N)}$ (calcu.) & $\chi^2$\\
\cline{2-6}
MD $\chi^2=5.96$
& 1.70$\pm$0.01 (E)
& 0.49
& 2.52$\pm$0.03 (G)
& 0.22
& 609\\
cf.
& 2.40$\pm$0.02 (E)
& 0.49
& 1.52$\pm$0.02 (E)
& 0.22
& 836\\
MD $\chi^2=6.16$
& 1.80$\pm$0.01 (E)
& 0.55
& 2.85$\pm$0.04 (G)
& 0.20
& 531 ($\ast 2$)\\
\hline
\end{tabular}
\end{table}


\begin{table}[htbp]
\vspace{-5mm}
\centering
\caption{\label{tab6}Results of the BEC data collected by the CMS~\cite{Khachatryan:2011hi} and analyzed by Eqs.~(\ref{eq10}), (\ref{eq11}), and (\ref{eq12}), where the BEC formulas are normalized with a constant factor and the long-range correlation is assumed as $(1+\varepsilon Q)$ (the labels ($\ast 1$ and $\ast 2$) indicate equivalence between the tabulated results by Eq.~(\ref{eq10}) and Eq.~(\ref{eq12}), respcetively).}
\vspace{1mm}
\begin{tabular}{c|ccccc}
\multicolumn{2}{l}{CMS\ 0.9 TeV}\\
\hline
& $R$ [fm] & $\lambda$ (free) & --- & --- & $\chi^2/$ndf\\
\cline{2-6}
CF$_{\rm I}$
& 1.56$\pm$0.02 (E)
& 0.62$\pm$0.01
& ---
& ---
& 487/194\\
& 0.87$\pm$0.01 (G)
& 0.30$\pm$0.00
& ---
& ---
& 1157/194\\
\hline
& $R_1^{\rm (II)}$ [fm] & $\lambda_1^{\rm (II)}$ (free) & $R_2^{\rm (II)}$ [fm] & $\lambda_2^{\rm (II)}$ (free) & $\chi^2$\\
\cline{2-6}
CF$_{\rm II}$
& 3.37$\pm$0.19 (E)
& 0.62$\pm$0.01
& 0.80$\pm$0.04 (G)
& 0.14$\pm$0.01
& 356\\
& 2.06$\pm$0.07 (G)
& 0.38$\pm$0.02
& 0.65$\pm$0.01 (G)
& 0.17$\pm$0.01
& 384 ($\ast 1$)\\
\hline
BEC$_{\rm (T\mathchar`-N)}$ & $R_1^{\rm (T\mathchar`-N)}$ [fm] & $\lambda_1^{\rm (T\mathchar`-N)}$ (calcu.) & $R_2^{\rm (T\mathchar`-N)}$ [fm] & $\lambda_2^{\rm (T\mathchar`-N)}$ (calcu.) & $\chi^2$\\
\cline{2-6}
MD $\chi^2 = 4.29$
& 2.02$\pm$0.02 (G)
& 0.46
& 0.61$\pm$0.01 (G)
& 0.15
& 429 ($\ast 1$)\\
MD $\chi^2 = 4.94$
& 2.06$\pm$0.02 (G)
& 0.45
& 0.62$\pm$0.01 (G)
& 0.16
& 422 ($\ast 1$)\\
cf.
& 1.29$\pm$0.01 (E)
& 0.45
& 2.04$\pm$0.01 (G)
& 0.15
& 454\\
\hline
\multicolumn{2}{l}{CMS 7.0 TeV}\\
\hline
& $R$ [fm] & $\lambda$ (free) & --- & --- & $\chi^2/$ndf\\
\cline{2-6}
CF$_{\rm I}$
& 1.89$\pm$0.02 (E)
& 0.62$\pm$0.01
& ---
& ---
& 738/194\\
& 1.03$\pm$0.01 (G)
& 0.29$\pm$0.00
& ---
& ---
& 1776/194\\
\hline
& $R_1^{\rm (II)}$ [fm] & $\lambda_1^{\rm (II)}$ (free) & $R_2^{\rm (II)}$ [fm] & $\lambda_2^{\rm (II)}$ (free) & $\chi^2$\\
\cline{2-6}
CF$_{\rm II}$
& 3.88$\pm$0.18 (E)  
& 0.84$\pm$0.03
& 0.71$\pm$0.01 (G)
& 0.12$\pm$0.01
& 540 ($\ast 2$)\\
& 2.39$\pm$0.07 (G)
& 0.40$\pm$0.01
& 0.76$\pm$0.01 (G)
& 0.16$\pm$0.00
& 600\\
\hline
BEC$_{\rm (T\mathchar`-N)}$ & $R_1^{\rm (T\mathchar`-N)}$ [fm] & $\lambda_1^{\rm (T\mathchar`-N)}$ (calcu.) & $R_2^{\rm (T\mathchar`-N)}$ [fm] & $\lambda_2^{\rm (T\mathchar`-N)}$ (calcu.) & $\chi^2$\\
\cline{2-6}
MD $\chi^2 = 2.27$
& 3.41$\pm$0.03 (E)
& 0.71
& 0.70$\pm$ 0.01 (G)
& 0.12
& 559 ($\ast 2$)\\
& 2.07$\pm$0.01 (E)
& 0.71
& 12.70$\pm$ 2.35 (E)
& 0.11
& 817\\
\hline
\end{tabular}
\end{table}


\section{\label{sec4}Concluding remarks and discussions}
Observing the results of Table~\ref{tab1} and Appendix~\ref{secA}, the T-NBD appears to adequately describe the MD at LHC energies. The MD data are contributed by three processes, ND, SD, and DD. The total probability distribution $P_{\rm tot}$ is expressed by the T-NBD (see Table~\ref{tab1} and Appendix~\ref{secA}).

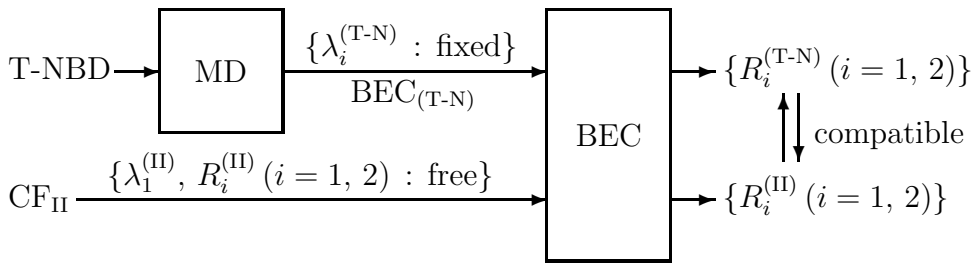
\begin{figure}[htbp]
\centering
\setlength\unitlength{1mm}
\begin{picture}(130,33)(0,0)
\thicklines
\put(0,24){T-NBD}
\put(0,7){CF$_{\rm II}$}
\put(14,25){\vector(1,0){6}}
\put(36,25){\vector(1,0){35}}
\put(87,25){\vector(1,0){6}}
\put(9,8){\vector(1,0){62}}
\put(87,8){\vector(1,0){6}}
\put(102,13){\vector(0,1){9}}
\put(104,22){\vector(0,-1){9}}
\put(20,17){\framebox(16,16){MD}}
\put(71,0){\framebox(16,33){BEC}}
\put(39,27){$\{\lambda_i^{\rm (T\mathchar`-N)}\,:\,{\rm fixed}\}$}
\put(45,21){BEC$_{\rm (T\mathchar`-N)}$}
\put(94,24){$\{R_i^{\rm (T\mathchar`-N)}\,(i=1,\,2)\}$}
\put(13,10){$\{\lambda_1^{\rm (II)},\,R_i^{\rm (II)}\,(i=1,\,2)\,:\,{\rm free}\}$}
\put(94,7){$\{R_i^{\rm (II)}\,(i=1,\,2)\}$}
\put(106,15.5){compatible}
\end{picture}
\caption{\label{fig3}Workflows of the T-NBD and CF$_{\rm II}$ computations. The interaction ranges of both computations are comparable.}
\end{figure}

Figure \ref{fig3} compares the workflows of the T-NBD and CF$_{\rm II}$ computations. The estimated interaction ranges and geometrical combinations are comparable between the two approaches.


The main achievements of the study are summarized below.

\paragraph{C1)} The values in Table \ref{tab2} are estimated after initializing the MD($P(n)$) values with random variables in the MINUIT application. Our calculations consider the lack of $P(0)$ and $P(1)$ in the ATLAS collaboration and the exclusion problem on $P(0)$ in the CMS collaboration~\cite{Zborovsky:2013tla}. Renormalization is the necessary step in the application of Eq.~(\ref{eq3}).

\paragraph{C2)} Utilizing the $\lambda_i^{\rm (T\mathchar`-N)}$ in Table~\ref{tab4}, we obtain the BEC data by Eq.~(\ref{eq10}). The $\chi^2$ values at 7 TeV are higher in CF$_{\rm I}$ than those in BEC$_{\rm (T\mathchar`-N)}$ and CF$_{\rm II}$ (see also point 3), probably because the MD($P(n)$) at the LHC is governed by stochastic effects~\cite{Biyajima:1983qu,Biyajima:1984ay}.

\paragraph{C3)} Using the CF$_{\rm II}$ values calculated by Eq.~(\ref{eq12}), we estimate the numerical values of the four-parameter set \{$R_i^{\rm (II)}$, $\lambda_1^{\rm (II)}$, $i=1,\,2$\}. The results are presented in Table~\ref{tab7}. The $R_1$ and $R_2$ values estimated by CF$_{\rm II}$ and BEC$_{\rm (T\mathchar`-N)}$ are satisfactorily similar. Table~\ref{tab8} summarizes the two degrees of coherence for the results indicated by ($\ast 1$ and $\ast 2$) in Tables \ref{tab5} and \ref{tab6}. Despite the large error bars in $\lambda_i^{\rm (T\mathchar`-N)}$'s, the $\lambda_i^{\rm (II)}$ and $\lambda_i^{\rm (T\mathchar`-N)}$ values are reasonably coincident, possibly reflecting the common stochastic properties of the MD and BEC ensembles, which are both described by the T-NBD.


\begin{table}[htbp]
\centering
\caption{\label{tab7}Comparison of $R_1$ and $R_2$ values marked with ($\ast 1$ and $\ast 2$) in Tables~\ref{tab5} and \ref{tab6}, and $\chi^2$ of the comparison.}
\vspace{1mm}
\begin{tabular}{c|c|ccc}
\hline
$\sqrt s$ [TeV] & formula & $R_1$ [fm] & $R_2$ [fm] & $\chi^2$\\
\hline
ATLAS
& CF$_{\rm II}$
& 2.82$\pm$0.28(G)
& 0.87$\pm$0.03 (G)
& 79.8\\
0.9
& BEC$_{\rm (T\mathchar`-N)}$
& 2.55$\pm$0.10 (G)
& 0.85$\pm$0.02 (G)
& 81.1\\
\hline
\lw{7.0}
& CF$_{\rm II}$
& 1.85$\pm$0.02 (E)
& 3.51$\pm$0.12 (G)
& 466\\
& BEC$_{\rm (T\mathchar`-N)}$
& 1.80$\pm$0.01 (E)
& 2.85$\pm$0.04 (G)
& 531\\
\hline
\hline
CMS
& CF$_{\rm II}$
& 2.06$\pm$0.07 (G)
& 0.65$\pm$0.01 (G)
& 384\\
0.9
& BEC$_{\rm (T\mathchar`-N)}$
& 2.06$\pm$0.02 (G)
& 0.62$\pm$0.01 (G)
& 422\\
\hline
\lw{7.0}
& CF$_{\rm II}$
& 3.88$\pm$0.18 (E)  
& 0.71$\pm$0.01 (G)
& 540\\
& BEC$_{\rm (T\mathchar`-N)}$
& 3.41$\pm$0.03 (E)
& 0.70$\pm$ 0.01 (G)
& 559\\
\hline
\end{tabular}
\end{table}


\begin{table}[htbp]
\vspace{-5mm}
\centering
\caption{\label{tab8}Comparisons of the two degrees of coherence for the results marked with ($\ast 1$ and $\ast 2$) in Tables~\ref{tab5} and \ref{tab6}.}
\vspace{1mm}
\begin{tabular}{c|c|cc}
\hline
$\sqrt s$ [TeV] & formulas & $\lambda_1$ & $\lambda_2$\\
\hline
ATLAS
& CF$_{\rm II}$
& 0.47$\pm$0.07
& 0.26$\pm$0.02\\
0.9 & BEC$_{\rm (T\mathchar`-N)}$
& 0.39$\pm$0.21
& 0.24$\pm$0.10\\
\hline
\lw{7.0}
& CF$_{\rm II}$
& 0.59$\pm$0.01
& 0.28$\pm$0.01\\
& BEC$_{\rm (T\mathchar`-N)}$
& 0.55$\pm$0.15
& 0.20$\pm$0.05\\
\hline
\hline
CMS
& CF$_{\rm II}$
& 0.38$\pm$0.02
& 0.17$\pm$0.01\\
0.9 & BEC$_{\rm (T\mathchar`-N)}$
& 0.45$\pm$0.25
& 0.16$\pm$0.11\\
\hline
\lw{7.0}
& CF$_{\rm II}$
& 0.84$\pm$0.03
& 0.12$\pm$0.01\\
& BEC$_{\rm (T\mathchar`-N)}$
& 0.71$\pm$0.30
& 0.12$\pm$0.09\\
\hline
\end{tabular}
\end{table}

\paragraph{C4)} Interesting interrelations are found between the results of BEC$_{\rm (T\mathchar`-N)}$ and CF$_{\rm II}$ marked with ($\ast 1$ and $\ast 2$) in Tables \ref{tab5} and \ref{tab6}. The BEC$_{\rm (T\mathchar`-N)}$ and CF$_{\rm II}$ may both reasonably describe the BEC at the LHC. Moreover, the combination of $E_{\rm BEC}^2$'s at 0.9 TeV satisfies the double-Gaussian distribution (G$+$G), whereas those at 7 TeV are combined exponential and Gaussian distribution (E$+$G). This finding implies different production mechanisms at 0.9 TeV and 7 TeV.

\paragraph{C5)} Possible correspondences are found among the KNO scaling, the MD, and the BEC (see Table~\ref{tab9}). These correspondences might be attributed to the violation of KNO scaling discovered in 1989 by the UA5 collaboration, that first proposed the D-NBD. The KNO scaling based on the T-NBD at LHC energies is calculated in Appendix~\ref{secB}.

\begin{figure}[htbp]
  \centering
  \includegraphics[width=0.60\columnwidth]{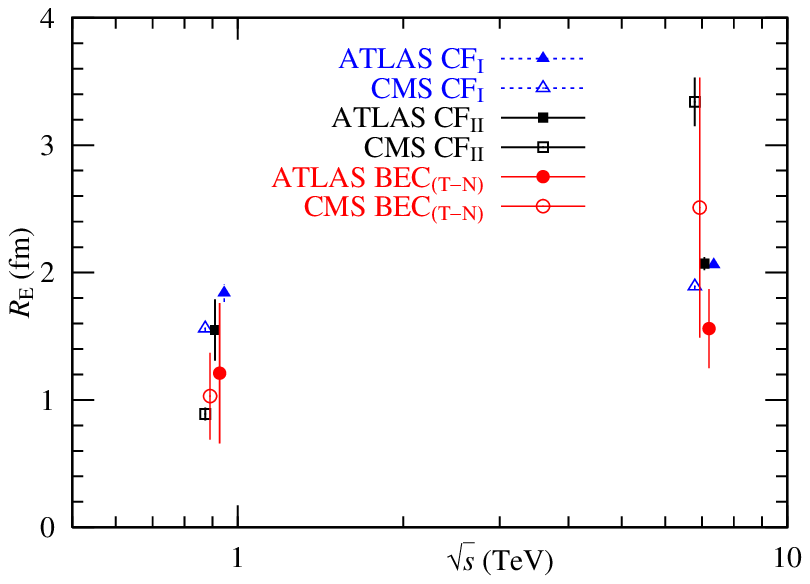}
  \caption{\label{fig4} Effective ranges calculated by Eq.~(\ref{eq13}). The $R$'s estimated by CF$_{\rm I}$ (E) are shown for comparison.}
\end{figure}

\paragraph{C6)} Taking into account the $\lambda_i^{\rm (II)}$'s as weight factors, the effective interaction ranges can be estimated as, 
\begin{eqnarray}
R_{\rm E} = R_1\times \lambda_1 + R_2\times \lambda_2.
\label{eq13}
\end{eqnarray}
The estimated effective interaction ranges are displayed in Table~\ref{tab10} and Fig.~\ref{fig4}. The $R_{\rm E}$ values appear reasonable because they are larger at the higher colliding energy (7.0 TeV) than at the lower energy (0.9 TeV).


\begin{table}[htbp]
\centering
\caption{\label{tab9}Correspondences among KNO scaling, MD, and BEC~\cite{Biyajima:2018abe,Mizoguchi:2019cra} (see also~\cite{Biyajima:1982un,Mizoguchi:2010vc}).}
\vspace{1mm}
\begin{tabular}{|c|c|c|}
\hline
KNO scaling & MD & BEC\\
\hline
existence. & Single NBD & CF$_{\rm I}$\\
$\psi_k(z)=\dfrac{k(kz)^{(k-1)}e^{-kz}}{\Gamma(k)},$ 
& Eq.~(\ref{eq2})
& Eq.~(\ref{eq11})\\
where $z=n/\langle n\rangle$.\qquad\qquad\qquad
& 
& \\
\hline
violation I: & D-NBD & CF$_{\rm II} =$ Eq.~(\ref{eq12})\\
$\psi(z)=\displaystyle\sum_{i=1}^2 \frac{\alpha_i}{r_i}\psi_{k_i}(z_i)$\hspace{8mm}
& $P(n,\, \langle n\rangle)$\hspace{30mm}
& ${\rm BEC}_{\rm (D\mathchar`-N)}$\hspace*{19mm}\\
$=\dfrac{\alpha_1}{r_1}\psi_{k_1}(z_1)+\dfrac{\alpha_2}{r_2}\psi_{k_2}(z_2)$,
& $= \displaystyle\sum_{i=1}^2 \alpha_i P_{{\rm NBD}_i}(n,\,\langle n_i\rangle,\,k_i)$,
& $= 1.0 + \lambda_1^{\rm (D\mathchar`-N)} E_{\rm BEC_1}^2$\\
where $z_i=z/r_i$~\cite{Biyajima:2018abe}.\qquad\qquad
& $s = \displaystyle\sum_{i=1}^2 \alpha_i \langle n_i\rangle^2 = a_1+a_2$.
& \ri{$\qquad + \lambda_2^{\rm (D\mathchar`-N)} E_{\rm BEC_2}^2$,}\\
$\displaystyle\sum_{i=1}^2 \alpha_i = 1.0$
& $\displaystyle\sum_{i=1}^2 \alpha_ir_i = 1.0$
& \ri{where $\lambda_i^{\rm (D\mathchar`-N)} = (a_i/s)(2/k_i)$}\\
& 
& \riw{($i=1,\,2$). See Ref.~\cite{Mizoguchi:2019cra}\qquad}\\
\hline
violation II: & T-NBD & CF$_{\rm II} =$ Eq.~(\ref{eq12})\\
$\psi(z)=\displaystyle\sum_{i=1}^3 \frac{\alpha_i}{r_i}\psi_{k_i}(z_i)$ 
& Eq.~(\ref{eq1})
& ${\rm BEC}_{\rm (T\mathchar`-N)}=$ Eq.~(\ref{eq10}),\\
$\displaystyle\sum_{i=1}^3 \alpha_i = 1$ 
& $s = \displaystyle\sum_{i=1}^3 \alpha_i \langle n_i\rangle^2 = \displaystyle\sum_{i=1}^3 a_i$
& where $\lambda_i^{\rm (T\mathchar`-N)} = (a_i/s)(2/k_i)$\\
The third term shows \quad
& $\displaystyle\sum_{i=1}^3 \alpha_ir_i = 1.0$
& \riw{($i=1\sim 3$).\ \ Notice\ \ that\quad}\\
\ri{the contribution of the}
& 
& \riw{$\lambda_3^{\rm (T\mathchar`-N)} = \mathcal{O}(10^{-3})$.\qquad\qquad}\\
\ri{Poisson-like distribution.\ }
& 
& \\
\hline
\end{tabular}
\end{table}


\begin{table}[htbp]
\centering
\caption{\label{tab10}Effective ranges calculated by Eq.~(\ref{eq13}).}
\vspace{1mm}
\begin{tabular}{c|c|cc}
\hline
& \lw{formulas} & \multicolumn{2}{c}{$R_{\rm E}$ [fm]}\\
& & 0.9 TeV & 7 TeV\\
\hline
\lw{ATLAS}
& CF$_{\rm II}$
& 1.55$\pm$0.24
& 2.07$\pm$0.05\\
& BEC$_{\rm (T\mathchar`-N)}$ 
& 1.21$\pm$0.55
& 1.56$\pm$0.31\\
\hline
\lw{CMS}
& CF$_{\rm II}$
& 0.89$\pm$0.05
& 3.34$\pm$0.19\\
& BEC$_{\rm (T\mathchar`-N)}$ 
& 1.03$\pm$0.34
& 2.51$\pm$1.02\\
\hline
\end{tabular}
\end{table}


\paragraph{D1)} We must also elucidate the physical meanings of the three intrinsic parameters $k_i$ and weight factor $\alpha_i$. From the MC data in~\cite{ATLAS:2010mza,GrosseOetringhaus:2009kz,Navin:2010kk} and the results of Table~\ref{tab1}, we infer the following correspondences:
\begin{eqnarray*}
\left\{
\begin{array}{l}
\mbox{1) The first NBD weighted with }\alpha_1\leftrightarrow\mbox{ the main part of }\sigma_{\rm ND}. \\
\mbox{2) The second NBD weighted with }\alpha_2\leftrightarrow\mbox{ the main part of }\sigma_{\rm SD}\mbox{, and parts of }\sigma_{\rm ND}\mbox{ and }\sigma_{\rm DD}. \\
\mbox{3) The third NBD weighted with }\alpha_3\leftrightarrow\mbox{ the main part of }\sigma_{\rm DD}\mbox{, and a part of }\sigma_{\rm SD}. \\
\end{array}
\right.
\end{eqnarray*}
Here, $\sigma_{\rm ND}$, $\sigma_{\rm SD}$, and $\sigma_{\rm DD}$ are the cross sections of the ND, SD, and double-diffractive dissociation (DD), respectively.

\paragraph{D2)} In future work, we are planning the following improvements:

\noindent The large error bars of $\lambda_i^{\rm (T\mathchar`-N)}\ (i=1,\,2)$ must be reduced in future work. For this purpose we must improve the framework of the MD analysis.

\paragraph{D3)} To properly validate the present theoretical formulation~\cite{Mizoguchi:2019a}, we will analyze the MD and BEC ($2.0<\eta<4.5$) obtained by the LHCb collaboration.\\

\noindent
{\it Acknowledgments.} One of the authors (M.B.) would like to thank his colleagues at the Department of Physics of Shinshu University for their kindness. T.~Mizoguchi would like to acknowledge the funding provided by Pres. Y.~Hayashi.


\appendix

\section{\label{secA}Monte Carlo data at 7 TeV collected by the ATLAS collaboration and analyzed by Eqs.~(\ref{eq1})--(\ref{eq3})}
\begin{figure}[htbp]
  \centering
  \includegraphics[width=0.60\columnwidth]{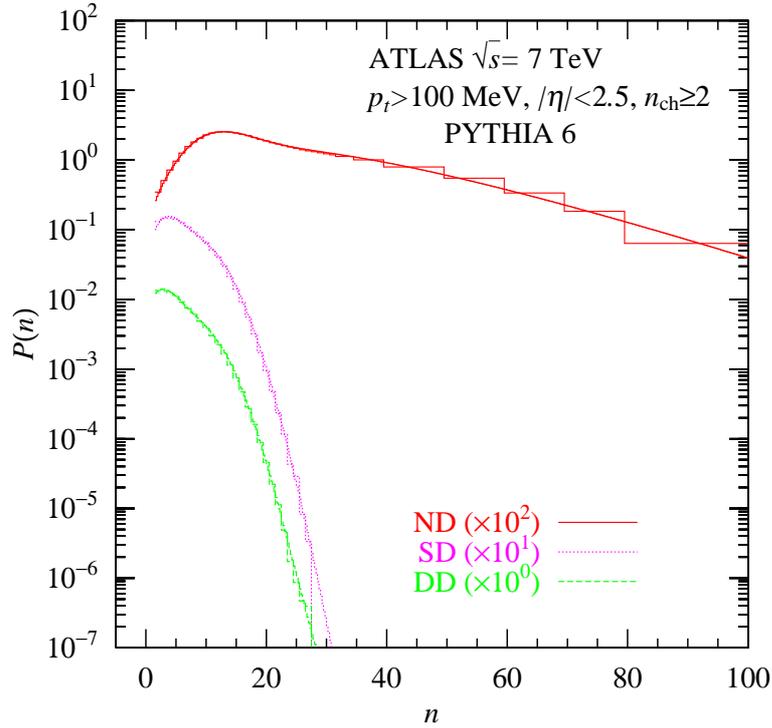}
  \caption{\label{fig5}Partial probability distributions $P_{\rm ND}$, $P_{\rm SD}$, and $P_{\rm DD}$~\cite{ATLAS:2010mza} obtained by Eq.~(\ref{eq2}) (see Table~\ref{tab11}).}
\end{figure}

\begin{figure}[htbp]
  \centering
  \includegraphics[width=0.60\columnwidth]{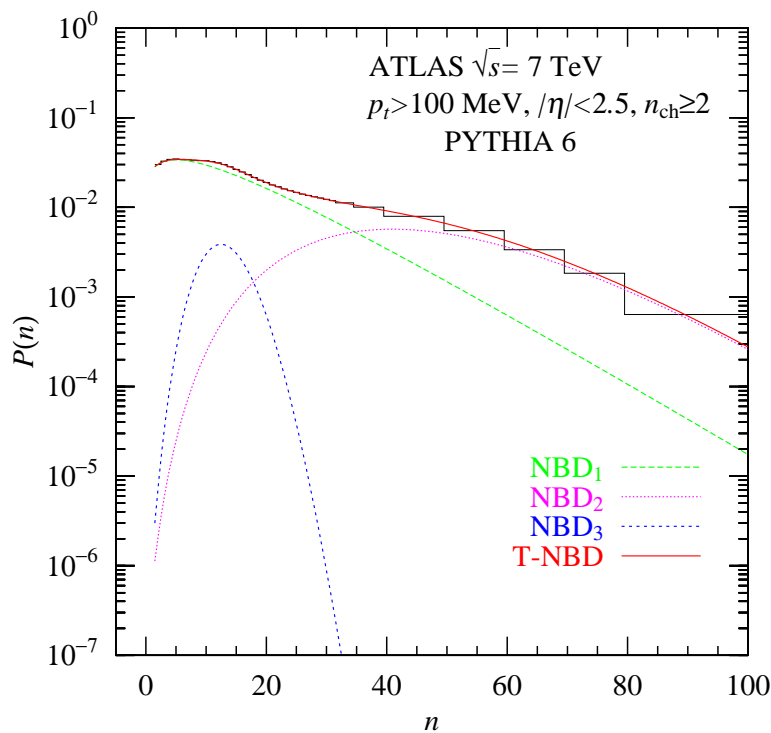}
  \caption{\label{fig6} Analysis of the total probability distribution $P_{\rm tot} = P_{\rm ND} + P_{\rm SD} + P_{\rm DD}$~\cite{ATLAS:2010mza} calculated by Eq.~(\ref{eq3}) (see Table~\ref{tab12}).}
\end{figure}

The MC data collected at 0.9 TeV and 7 TeV are presented in~\cite{ATLAS:2010mza}, which reported the MD results. The total MD is decomposed into three processes: ND, SD, and double-diffractive dissociation (DD), with probability distributions defined by $P_{\rm ND}$, $P_{\rm SD}$, and $P_{\rm DD}$, respectively. The total probability distribution is expressed as $P_{\rm tot} = P_{\rm ND} + P_{\rm SD} + P_{\rm DD}$. The partial and total probability distributions are plotted in Figs.~\ref{fig5} and \ref{fig6}, respectively. 

The NBD and D-NBD are calculated by Eqs.~(\ref{eq1}) and (\ref{eq2}), respectively, and the results are shown in Table~\ref{tab11}. Obviously, the single-NBD cannot describe the MD. The D-NBD probably constitutes three ensembles with different $\langle n\rangle$ and $k$ values: the first set with ($\langle n\rangle\cong 33$, $k = $2--4), the second set with ($\langle n\rangle = $4--13, $k = $4--16), and the third set with ($\langle n\rangle \cong 10$ and $k =$200--1000). These ensembles appear to reasonably validate the T-NBD framework in the analysis of MD at the LHC. The T-NBD is computed from the total MC data at 7 TeV by Eq.~(\ref{eq3}). The analysis results are shown in Table~\ref{tab12}. 

We also analyze the MC data at 0.9 TeV by PYTHIA 6, and at 7 TeV by PHOJET and PYTHIA 8. The results are similar to those in Tables~\ref{tab11} and \ref{tab12}.


\begin{table}[htbp]
\centering
\caption{\label{tab11}PYTHIA 6 analysis of MD at 7 TeV by the ATLAS collaboration calculated by Eqs.~(\ref{eq1}) and (\ref{eq2}). The magnitude of the error bars is assumed as 10\% of the data points. The $\chi^2$ values are markedly improved by D-NBD.}
\vspace{1mm}
\begin{tabular}{cc|cc|ccccc}
\hline
PYTHIA 6 & ATLAS & \multicolumn{2}{c}{single-NBD} & \multicolumn{5}{|c}{D-NBD}\\
7 TeV & ratio & $\langle n\rangle$, $k$ & $\chi^2$ & $i$ & $\alpha_i$ & $\langle n_i\rangle$ & $k_i$ & $\chi^2$\\
\hline
\lw{ND}
& \lw{0.787}
& $\langle n\rangle = 30.17\pm 0.36$
& \lw{150.4}
& 1
& 0.814
& 32.5
& 2.56
& \lw{6.84}\\
&
& $k = 2.53\pm 0.04$
&
& 2
& 0.186
& 13.2
& 16.85
& \\
\hline
\lw{SD}
& \lw{0.121}
& $\langle n\rangle = 7.08\pm 0.08$
& \lw{257.3}
& 1
& 0.716
& 4.63
& 7.56
& \lw{11.75}\\
&
& $k = 15.87\pm 0.80$
& 
& 2
& 0.284
& 10.1
& 206.7
& \\
\hline
\lw{DD}
& \lw{0.092}
& $\langle n\rangle = 6.15\pm 0.08$
& \lw{240.7}
& 1
& 0.798
& 4.14
& 4.34
& \lw{34.00}\\
&
& $k = 10.46\pm 0.50$
& 
& 2
& 0.202
& 9.88
& 1000.0
& \\
\hline
\end{tabular}
\end{table}


\begin{table}[htbp]
\centering
\caption{\label{tab12}Estimated parameters of T-NBD in the $P_{\rm tot} = P_{\rm ND} + P_{\rm SD} + P_{\rm DD}$ (calculated by Eq.~(\ref{eq3})). The magnitude of the error bars is assumed as 10\% of the data points.}
\vspace{1mm}
\begin{tabular}{cc|cc|cccc}
\hline
$i$ & $\alpha_i$ & $\langle n_i\rangle$ & $k_i$ & $\chi^2$\\
\hline
 1
& 0.711$\pm$0.131
& 15.37$\pm$3.62
& 1.56$\pm$0.31
& \\
 2
& 0.255$\pm$0.129
& 47.86$\pm$5.69
& 7.40$\pm$2.88
& 0.168\\
 3
& 0.035$\pm$0.023
& 12.99$\pm$2.05
& 1000.0$\pm$551.0
& \\
\hline
\end{tabular}
\end{table}


\section{\label{secB}An Enlarged KNO scaling function for LHC energies}
In this Appendix, we analyze the KNO scaling at LHC energies. Recall that the D-NBD was proposed to explain the KNO scaling violation found at the S$\bar{p}p$S energy ($\sqrt s = 546$ GeV)~\cite{Fuglesang:1989st}. The KNO scaling function in the framework of the T-NBD~\cite{Biyajima:2018abe} is given by
\begin{eqnarray}
  \psi_{\rm (T\mathchar`-N)}(z=n/\langle n\rangle,\,\alpha_i,\,k_i,\,r_i;\,i=1\sim 3) = \sum_{i=1}^3 \alpha_i \frac 1{r_i} \frac{k_i^{k_i}(z/r_i)^{k_i-1}}{\Gamma(k_i)}e^{-k_iz/r_i}.
\label{eq14}
\end{eqnarray}
After integratin of the KNO scaling variables $z$, the KNO scaling function becomes
\begin{eqnarray}
  \int_0^{\infty} dz\psi_{\rm (T\mathchar`-N)}(z,\,\alpha_i,\,k_i,\,r_i;\,i=1\sim 3) = \sum_{i=1}^3 \alpha_i = 1.0.
\label{eq15}
\end{eqnarray}
The violation of the KNO scaling can be understood studying the energy dependences of the parameters $\alpha_i$, $k_i$, and $r_i$. 

\begin{figure}[htbp]
  \centering
  \includegraphics[width=0.48\columnwidth]{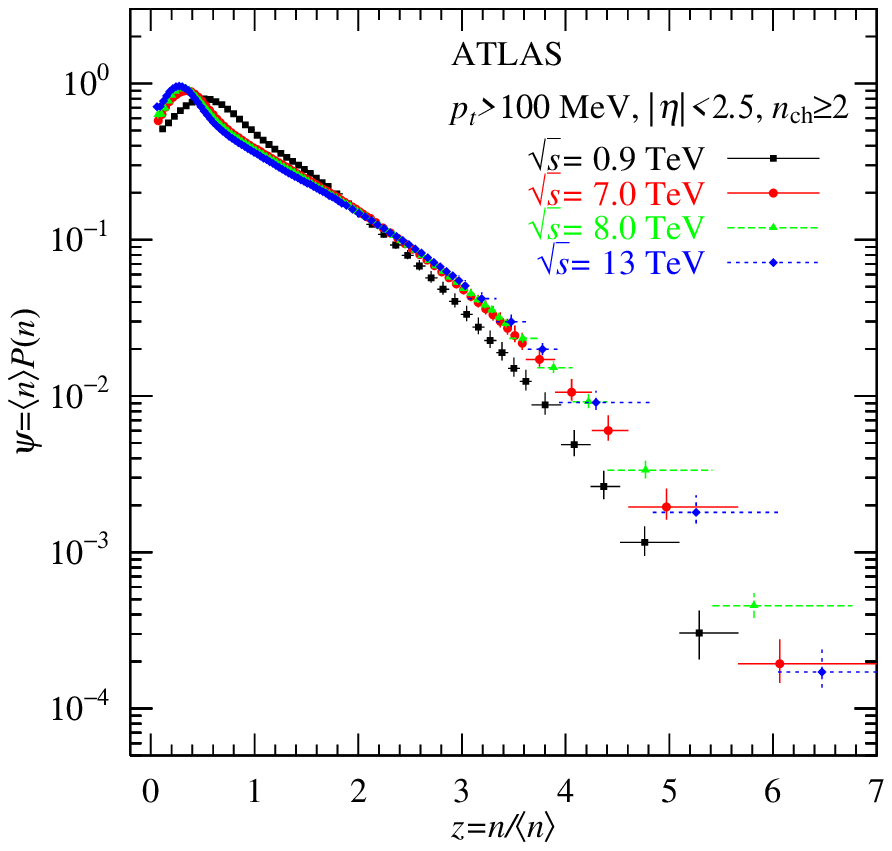}
  \includegraphics[width=0.48\columnwidth]{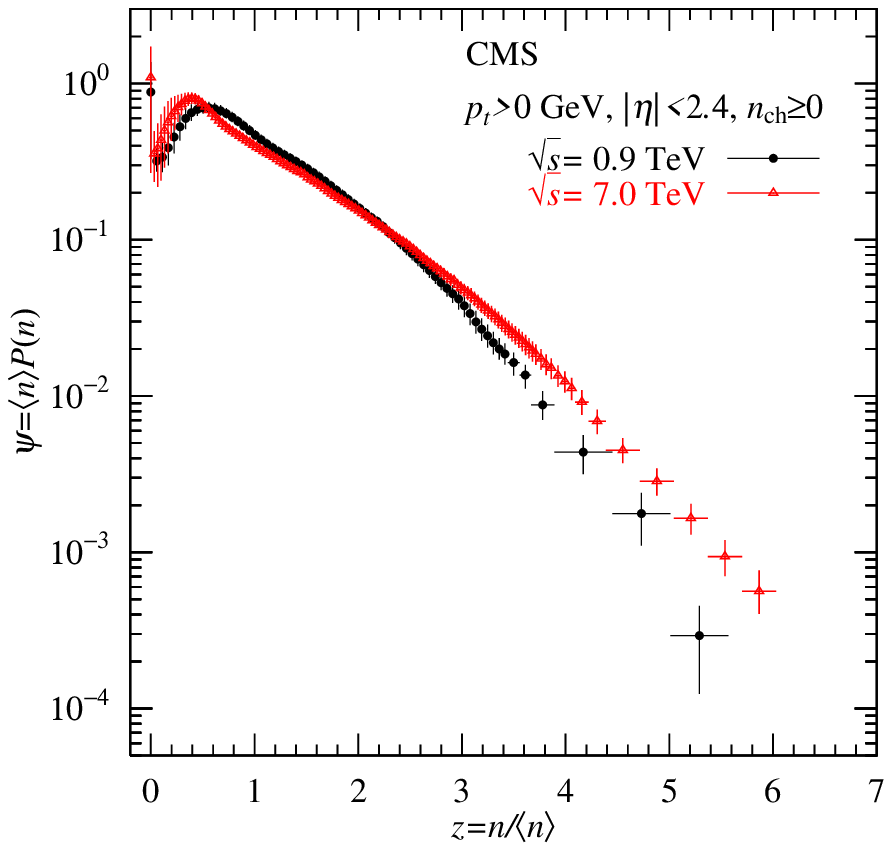}\\
  \includegraphics[width=0.48\columnwidth]{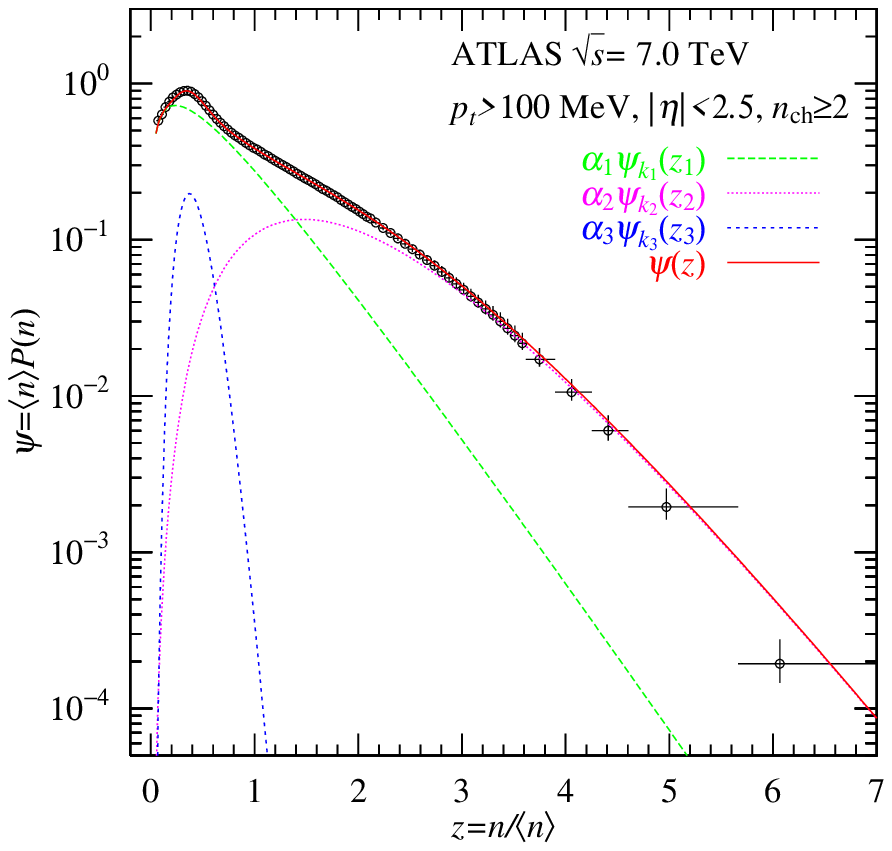}
  \includegraphics[width=0.48\columnwidth]{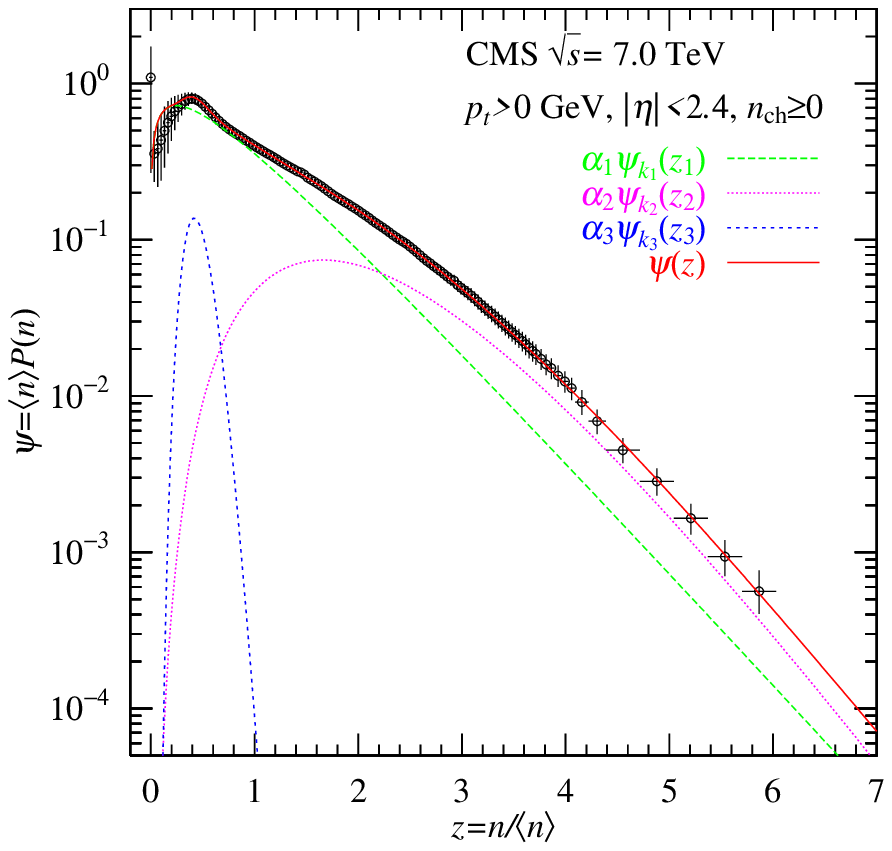}
  \caption{\label{fig7}Analysis of KNO data collected by the ATLAS and CMS collaborations. The KNO scaling functions were computed by Eq.~(\ref{eq14}).}
\end{figure}

The results of Eq.~(\ref{eq14}) are shown in Fig.~\ref{fig7} and Table~\ref{tab13}. The ratio $r_i=\langle n_i\rangle/\langle n\rangle$ should be large, because the average multiplicities $\langle n_i\rangle$ are requaired to be large as $\langle n\rangle$ itself; specifically, $r_i\ge 0.33$ ($\langle n_i\rangle \ge \langle n\rangle/3$). The KNO scaling functions at 0.9 TeV obtained by the ATLAS and CMS collaborations differed from from those at 7, 8, and 13 TeV (Fig~\ref{fig7}). The 0.9 TeV data collected by the CMS collaboration must be constrained by $k_2 > 5$ and $k_3 > 7$ because $z_3$ is large (800 or infinity) in the MD analysis. Our KNO scaling analysis also includes the $P(0)$ data at 0.9 and 7 TeV obtained by the CMS collaboration. As shown in Table~\ref{tab13}, the violation of KNO scaling occurs through the parameters $\alpha_i$ and $r_i=\langle n_i\rangle/\langle n\rangle$ ($i=1\sim 3$).


\begin{table}[htbp]
\centering
\caption{\label{tab13}Analysis (by Eq.~(\ref{eq14})) of KNO data at 0.9 and 7.0 TeV collected by the ATLAS and CMS collaboration.} 
\vspace{1mm}
\begin{tabular}{c|cccc}
\hline
& $i$ & $\alpha_i$ & $r_i$ & $k_i$\\
\hline
  ATLAS
& 1
& 0.761$\pm$0.134
& 0.84$\pm$0.10
& 1.79$\pm$0.08\\
  0.9 TeV
& 2
& 0.166$\pm$0.006
& 1.87$\pm$0.03
& 5.75$\pm$1.62\\
$\chi^2 = 25.6$
& 3
& 0.074$\pm$0.031
& 0.67$\pm$0.01
& 11.0$\pm$3.2\\
\hline
  ATLAS 
& 1
& 0.664$\pm$0.067
& 0.68$\pm$0.09
& 1.54$\pm$0.10\\
  7.0 TeV
& 2
& 0.275$\pm$0.011
& 1.91$\pm$0.02
& 4.37$\pm$0.54\\
$\chi^2 = 27.4$
& 3
& 0.061$\pm$0.011
& 0.41$\pm$0.01
& 10.0$\pm$1.8\\
\hline
ATLAS
& 1
& 0.692$\pm$0.069
& 0.73$\pm$0.11
& 1.50$\pm$0.10\\
8.0 TeV
& 2
& 0.235$\pm$0.003
& 2.00$\pm$0.01
& 4.61$\pm$0.61\\
$\chi^2 = 39.28$
& 3
& 0.073$\pm$0.015
& 0.38$\pm$0.01
& 7.77$\pm$1.34\\
\hline
ATLAS
& 1
& 0.751$\pm$0.037
& 0.81$\pm$0.06
& 1.27$\pm$0.04\\
13 TeV
& 2
& 0.168$\pm$0.050
& 2.18$\pm$0.09
& 4.88$\pm$0.49\\
$\chi^2 = 43.26$
& 3
& 0.081$\pm$0.007
& 0.35$\pm$0.01
& 7.33$\pm$0.66\\
\hline
  CMS
& 1
& 0.575$\pm$0.079
& 0.75$\pm$0.09
& 1.51$\pm$0.17\\
  0.9 TeV
& 2
& 0.286$\pm$0.050
& 1.66$\pm$0.04
& 5.0 (lower limit)\\
$\chi^2 = 18.80$
& 3
& 0.139$\pm$0.043
& 0.67$\pm$0.04
& 7.0 (lower limit)\\
\hline
  CMS
& 1
& 0.809$\pm$0.019
& 0.83$\pm$0.02
& 1.42$\pm$0.05\\
  7.0 TeV
& 2
& 0.153$\pm$0.009
& 2.06$\pm$0.05
& 5.18$\pm$0.33\\
$\chi^2 = 7.83$
& 2
& 0.038$\pm$0.020
& 0.45$\pm$0.04
& 15.2$\pm$10.9\\
\hline
\end{tabular}
\end{table}


\end{document}